\documentclass[twocolumn, prX, longbibliography,aps]{revtex4-1}

\usepackage{graphicx}
\usepackage{multirow}
\usepackage{dcolumn}
\usepackage{bm}
\usepackage[normalem]{ulem}
\usepackage{amsmath}
\usepackage{amsfonts}
\usepackage{amssymb}
\usepackage{color}
\usepackage{colordvi}
\usepackage{dcolumn}

\usepackage[colorlinks=true, breaklinks=true, linkcolor=blue,
urlcolor=blue, citecolor=blue]{hyperref}

\begin{document}

\title{Novel phases in twisted bilayer graphene at magic angles as a result of van Hove singularities and interactions}
\author {Yury Sherkunov and Joseph J. Betouras}
\email{J.Betouras@lboro.ac.uk or Y.Sherkunov@lboro.ac.uk}
\affiliation {Department of Physics and Centre for the Science of Materials, Loughborough University, Loughborough, LE11 3TU, United Kingdom}
\date{\today}

\begin{abstract}

The discovery of different phases as a result of correlations, especially  in low-dimensional materials, has been always an exciting and fundamental subject of research. Recent experiments on twisted bilayer graphene have revealed reentrant unconventional superconductivity as a function of doping as well as a Mott-like insulating phase when the two layers are twisted  with respect to each other at certain "magic" angles for doping corresponding to two particles per moire unit cell. In this work we propose a microscopic model  that takes into account interactions and the van Hove singularities in the density of states of the twisted bilayer graphene at doping corresponding to one particle ($\nu$ =1) per moir\'{e} unit cell and study how superconductivity emerges. We identify the possible symmetry of the order parameter as $s^{\pm}$, while if the inter-valley coupling is negligible the symmetry is $s^{++}$. In addition, we find and characterise  the insulating region of the system, as a region with a uniform charge instability where there is coexistence of the metallic and insulating phases.

 \end{abstract}

\maketitle

\section{Introduction.} 

Recent experiments on twisted bilayer graphene have revealed the importance of the effects of correlations and the development of unconventional superconductivity in these two-dimensional systems \cite{CaoMott18, CaoSC18}.  One fundamental ingredient of this physics is that by twisting the two layers in the bilayer system with respect to each other at precisely some desired angles, the layers hybridize such as to form flat bands near the Fermi level.  This in turn leads to Lifshitz transitions where the Fermi velocity goes to zero and the density of states (DOS) gets enhanced. Indeed van Hove singularities were observed in twisted bilayer graphene in an earlier work \cite{Andrei2009}. In a broader sense, the system is then susceptible to the formation of different phases as a result of the interactions.  Recent examples of the role of Lifshitz transitions in correlated systems  include ferromagnetic superconductors \cite{Sherkunov_Chubukov_Betouras_2018}, pnictides \cite{Liu_2010}, cobaltates \cite{Slizovskiy_Chubukov_Betouras_2015}.

The novel phases that have been discovered in twisted bilayer graphene is superconductivity and a Mott-like insulator behavior in the case of hole-doped bilayer graphene with filling factor clearly at $\nu=2$, corresponding to two particles per unit cell of the moir\'{e} pattern.  At $\nu=1$ there is still a debate on the nature of the insulating-like state \cite{Phillips18}. Deeper understanding of both phases and their relation is of fundamental importance; this has been the subject of intense studies since the discovery of high-temperature superconductors \cite{LeeRMP2006} and is  attracting a surge of interest in relation to twisted bilayer graphene \cite{Cenke18, Po18,Phillips18, Dodaro18,Isobe18,Liu18, Zhuang18}. 

In this work, we study the effects of correlations, taking into account the singularities in the DOS and provide an explanation of the phase diagram in the temperature-density plane for the case of singe occupation of the moir\'{e} unit cell, $\nu=1$. The physics at $\nu=1$ should be distinct from the one associated with higher filling factors. The reason is that according to the experimental results of  \cite{CaoMott18, CaoSC18}, the renormalized Fermi velocity in the vicinity of the magic angle  is of order of $v^*=4\times 10^4m/s$, which is $25$ times smaller than the one of a single layer graphene. This puts the position of the van Hove singularities in question at $\epsilon_0=0.25meV$, which corresponds to the filling factor $\nu=1$ per moir\'{e} unit cell \cite{Liu18}. In addition, there is no nesting at that filling, contrary to the structure of the singularities associated with higher filling factors.

We find that the system shows  reentrant behavior of the superconductivity, for which  we predict that the order parameter symmetry is $s^{\pm}$ or $s^{++}$, therefore different that the one predicted for single layer graphene \cite{Uchoa07, Nandkishore12} and we provide the reason for that. We also find  a phase of  uniform charge instability (UCI), with coexistence of insulating and metallic regions.

The structure of the paper is that in the next section we provide a discussion on the effective Hamiltonian, in Sec. III there is a discussion on the polarisation operators and the structure of RG equations, in Sec IV the results of RG analysis are presented and, finally, in Sec. V we discuss the results in the context of the experimental work.

\section{Effective Hamiltonian.} 

\begin{figure}[h]
\includegraphics[width=0.45\textwidth]{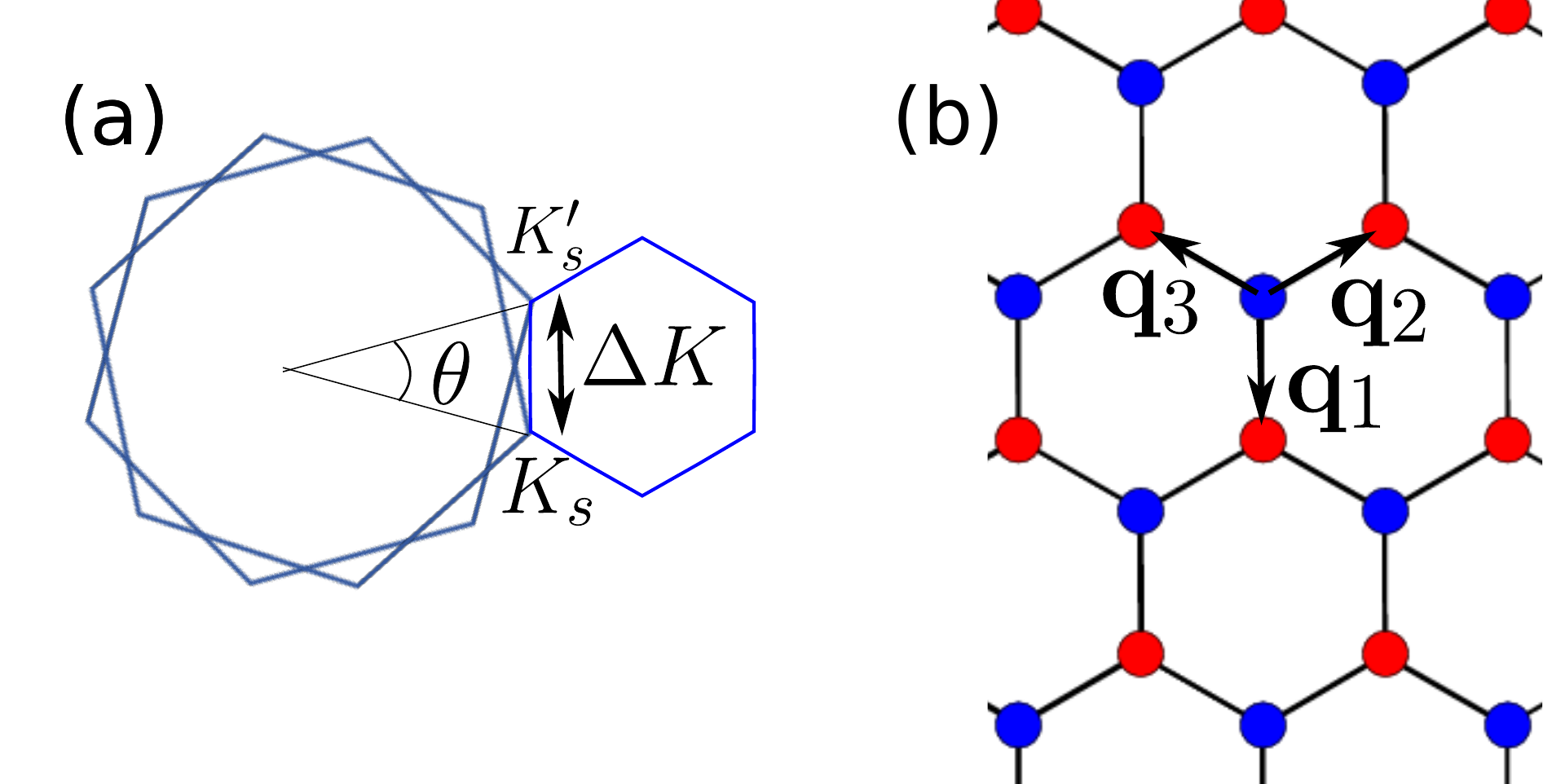}
\caption{ Momentum-space geometry of a  bilayer graphene twisted by angle $\theta$.  (a) The Brillouin zones of the two graphene layers are shown on the left and the mini Brillouin zones of the twisted bilayer graphene constructed from the difference, $\Delta \mathbf K=\mathbf K_s-\mathbf K_s'$,  between the two vectors of the Dirac points, $\mathbf K_s$ and $\mathbf K_s'$, of the two layers are shown on the right. At small angles, $\Delta K=|\mathbf K_s| \theta$. (b) The three
equivalent Dirac points in the first mini Brillouin zone result in three distinct hopping processes described by vectors $\mathbf q_1=\Delta K(0,-1)$, $\mathbf q_2=\Delta K(\sqrt 3/2,1/2)$, and $\mathbf q_3=\Delta K(-\sqrt 3/2,1/2)$ generating a honeycomb lattice in the k-space. }
\label{Fig1}\end{figure}

A twisted bilayer graphene has a moir\'{e} superlattice pattern, which is reciprocal to a hexagonal mini Brillouin zone, with side, $\Delta K=K\theta$, equal to the difference between two $K$-vectors of the twisted layers as shown in Fig. \ref{Fig1} (a). The electronic spectrum of twisted bilayer graphene has been thoroughly studied \cite{dosSantos2007, Mele2010a, Mele2010b}. A continuous low-energy theory developed in Ref.\cite{MacDonald11}, with interlayer tunnelling only between the Dirac points of the mini Brillouin zone parametrised by vectors $\mathbf q_{1,2,3}$ generating a $k$-space honeycomb lattice corresponding to repeated hopping, as shown in Fig. \ref{Fig1} (b). We adopt this theory to study the low energy spectrum and derive the effective Hamiltonian as a starting point.

In the simplest limit in which the momentum space lattice is truncated to the first honeycomb shell, the system can be described by the following Hamiltonian \cite{MacDonald11}:

\begin{equation}
H=\begin{bmatrix}
    h_k       & T_1 & T_2 & T_3 \\
    T_1^{\dagger}       & h_{k_1} &0 & 0 \\
    T_2^{\dagger}       & 0 & h_{k_2} &0\\
    T_3^{\dagger} & 0 &0 &h_{k_3}
\end{bmatrix},\label{Hamiltonian}
\end{equation}
where  $h_{k_i}=-v\bm{\sigma}^*\cdot \mathbf k_i$ is the Dirac Hamiltonian in the vicinity of one of the four Dirac points of the first honeycomb shell  connected by the vectors $\mathbf q_i$, as shown in Fig. \ref{Fig1} (b), $\mathbf k_i=\mathbf k+\mathbf q_i$,  $v$ is the bare Fermi velocity, and $\bm{\sigma}$ is a vector of Pauli matrices. The tunnelling matrix elements are given by: 
\begin{equation}
T_i=w\begin{bmatrix}
    e^{-i\phi_i}      & 1 \\
    e^{i\phi_i}       & e^{-i\phi_i} 
\end{bmatrix},\label{T}
\end{equation}
where $w$ is the hopping energy $\phi_1=0$, $\phi_2=2\pi/3$, and $\phi_3=-2\pi/3$.

In first order perturbation theory in $k$, the effective low-energy Hamiltonian is written as \cite{MacDonald11}
\begin{equation}
H_1=-v^*\bm{\sigma}^*\cdot \mathbf k, \label{H1}
\end{equation}
where $v^*$ is the renormalised Fermi velocity given by $v^*=v\frac{1-3\alpha^2}{1+6\alpha^2}$, where $\alpha=w/(v\Delta K)$. For $\alpha=\alpha_0=1/\sqrt 3$, $v^*$ vanishes, leading to the  flattening of the  low-energy bands. This happens at one of the "magic" twist angles. 
Then using second order perturbation theory in $k$ the next order of the effective Hamiltonian reads
\begin{eqnarray}
H_2&=&\frac{3\alpha^2v}{(1+6\alpha^2)\Delta K k}\nonumber\\
&\times& \begin{bmatrix}
    k_x(k_x^2-3k_y^2)      & (k_x+ik_y)(3k_x^2-k_y^2)k_y/k \\
    (k_x-ik_y)(3k_x^2-k_y^2)k_y/k       &- k_x(k_x^2-3k_y^2)  
\end{bmatrix}.\nonumber
\end{eqnarray}
Diagonalising the Hamiltonian $H=H_1+H_2$, we find the eigenvalues, which coincide with the ones obtained from a phenomenological $k$-expansion of the low energy Hamiltonian \cite{CaoMott18}, 
\begin{equation}
\epsilon=\pm \sqrt{v^{*2}k^2-\frac{v^*k^3\sin(3\beta)}{m}+\frac{k^4}{4m^2}},\label{EV}
\end{equation}
where $\beta$ is the angle between $\mathbf k$ and $k_x$, however, our approach enabled us to identify the parameter $m$ as $m=\frac{(1+6\alpha^2)\Delta K}{6\alpha^2v}$. 

\begin{figure}[h]
\includegraphics[width=0.45\textwidth]{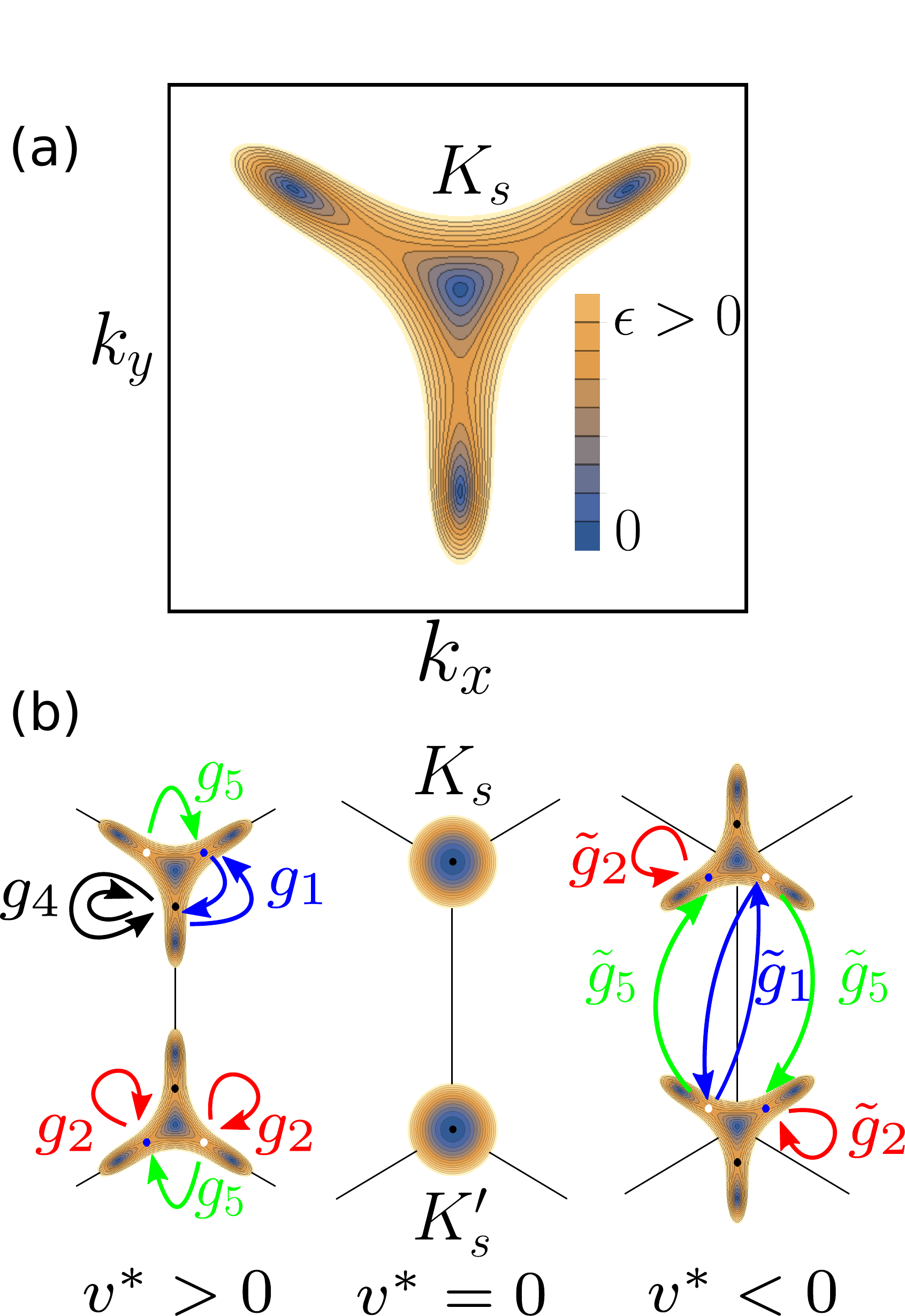}
\caption{(a) Low-energy electronic dispersion in the vicinity of a $K-$point of TBG for $\epsilon>0$ and $v^*>0$. Contours show the Fermi-surface family, with Lifshitz transition occurring when the chemical potential crosses the saddle points. (b) Pictorial representation of the scattering processes shown on the plot of  energy dispersion for two non-equivalent Dirac points of TBG for $v^*>0$ (left), $v^*=0$ (middle), and $v^*<0$ (right). Dots of the same colour indicate the saddle points with identical energy dispersion (conjugate points).    }
\label{Fig2}\end{figure}

The energy spectrum (\ref{EV}) has three saddle points, as shown in Fig. \ref{Fig2} (a),  located at $\{k_{xsp},k_{ysp}\}$ with $|k_{sp}|=m|v^*|$. In the vicinity of the saddle points, the energy can be expanded as
\begin{eqnarray}
\epsilon_1&=&\frac{1}{2m}(9\delta k_x^2-\delta k_y^2)\nonumber,\\
\epsilon_{2/3}&=&\frac{1}{4m}(3\delta k_x^2\mp 10\sqrt 3 \delta k_x\delta k_y+13 \delta k_y^2)\label{e1},
\end{eqnarray}
where $\delta k_j=k_j-k_{jsp}$. The values of the energies are counted as the difference from their values at the saddle points,  $\epsilon_0=mv^{*2}/2$, which can be absorbed into the chemical potential.  The azimuthal positions of the saddle points are determined by the valley index and the sign of $v^*$, so that for the valley $K_s$ and $v^*>0$,  $\beta_{sp}=\pi/6,\; 5\pi/6,\;3\pi/2$. For $K_s'$ valley or $v^*<0$, the positions can be obtained  by inversion, $\mathbf k_{sp} \rightarrow -\mathbf k_{sp}$. Thus, for each saddle point in valley $K_s$, there is a conjugate saddle point with identical energy in valley $K_s'$, as shown in Fig. \ref{Fig2} (b).   The saddle points shift towards the point $k=0$, as $|v^*|$ decreases, where they merge at $v^*=0$ to form a minimum, as shown in the middle panel of Fig. \ref{Fig2} (b), in contrast to Ref. \cite{Shtyk17}, where by controlling the gate voltage, the three saddle points merge to form a monkey saddle point leading to a power-law singularity in the DOS.   

\section{Polarisation operators and RG analysis.} 

The presence of the saddle points leads to a logarithmically divergent van Hove singularity in the DOS per spin, per saddle point,
\begin{equation}
\nu(\epsilon)=\nu_0\ln\left |\frac{\Lambda}{\epsilon}\right |,\label{Dos}
\end{equation}
where $\nu_0=\frac{8m}{9 \sqrt 3\Delta K^2}$, and $\Lambda$ is the usual ultraviolet cut-off. Note that at $v^*=0$, the saddle points merge into a minimum and the DOS becomes constant.

Due to the logarithmic divergence of the DOS and the polarisation operators (shown below), the renormalisation group (RG) theory is the major tool at work. We follow the standard procedure developed in \cite{Furukawa98, Rice10} that has been also used in monolayer graphene doped up to the M-points \cite{Nandkishore12}. In this procedure, the fermions that are taken into account are those that live in patches around each of the six saddle points with logarithmically divergent  DOS. 

It is worth emphasising again that the renormalized Fermi velocity in the vicinity of the magic angle  is of order of $v^*=4\times 10^4m/s$ \cite{CaoMott18, CaoSC18}, thus $25$ times smaller than the one of a single layer graphene. As a result, the position of the van Hove singularities is at $\epsilon_0=0.25meV$, corresponding to the filling factor $\nu=1$.  We will adopt this value for our calculations. Moreover, there is no nesting in this case.

The screening of the Coulomb interaction, $ U(k)=2\pi e^2/(k a^2)$, due to high-energy states can be estimated  using random phase approximation (RPA) \cite{Nandkishore12}
\begin{equation}
U(k)=\frac{2\pi e^2/(ka^2)}{1+2\pi e^2N\Pi_{\Lambda_0}/(ka^2)},\label{U}
\end{equation}
where $a$ is the carbon-carbon distance in a monolayer graphene, $N=12$ is the number of fermionic flavours, and $\Pi_{\Lambda_0}$ is the polarazation operator taking into account all the states between some ultraviolet cutoff, $\Lambda_0$, and the band-width, $W$. For large $k$, it can be estimated as \cite{Nandkishore12} $\Pi_{\Lambda_0}(k)\approx\nu(\Lambda_0)=\nu_0\ln\left|\Lambda/\Lambda_0\right|$. This allows us to rewrite Eq. (\ref {U}) as
\begin{equation}
\nu_0U(k)=\frac{Z(k)}{1+NZ(k)},\label{UZ}
\end{equation}
with $Z(k)=2\pi e^2\nu_0/(ka^2)$, where we assumed $\Lambda_0=\Lambda$. The two-particle scattering between patches are determined by the following characteristic momenta: $k_1\approx m v^*/3$ for intra-patch scattering, $k_2\approx m v^*$ for interpatch scattering within the same valley, and $k_3\approx \Delta K$ for inter-valley scattering, for which we estimate $Z(k_1)\approx \frac{10\sqrt 3 v}{\pi v^*\theta^2}$, $Z(k_2)=Z(k_1)/3$, and $Z_3\approx \frac{5 \sqrt 3}{\pi\theta^2}$. Here we used $e^2/v\approx 10/3$, for $a\approx 1\AA$ and $v\approx 10^6m/s$. For $v^*\ll v$ and $\theta\approx 1^{\circ}$, $Z(k_i)N\gg 1$ for $i=1,2,3$, and the Coulomb potential is completely screened, so that $\nu_0U(k_1)=\nu_0U(k_2)=\nu_0U(k_3)=1/N\approx 0.083$. This allows us to assume that the coupling is valley independent.

\begin{figure}[h]
\includegraphics[width=0.45\textwidth]{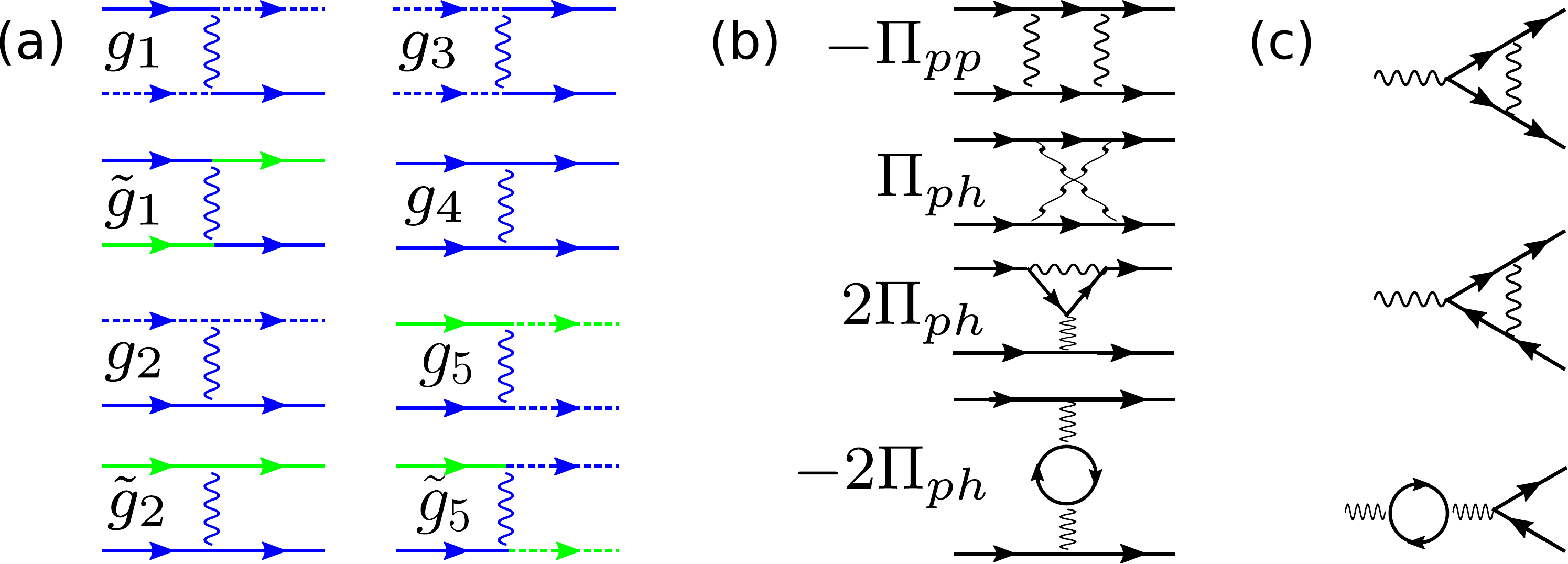}
\caption{(a) Feynman diagrams representing two-particle scattering processes between different patches (see Fig.\ref{Fig2} b). Solid and dashed lines represent fermions on different patches, with different colours marking  two patches of different valleys with identical energies (conjugate points). Wavy lines represent interactions.  Note that the Umklapp  scattering, $g_3$, is forbidden due to momentum conservation. (b)-(c) One-loop contributions to the renormalisation of the interaction constants (b) and test vertices (c).}
\label{Fig3}\end{figure}

The two-particle scattering between patches is described by the eight distinct interactions in the low-energy theory depicted in Fig. \ref{Fig3} (a) and visualised in Fig. \ref{Fig2} (b). The system is described by the low-energy Lagrangian:
\begin{eqnarray}
\mathcal{L}&=&\sum_{\alpha,\sigma}\psi_{\alpha\sigma}^{\dagger}(\partial_{\tau}-\epsilon_k+\mu)\psi_{\alpha\sigma}\nonumber\\
&-&\frac{1}{2}\sum_{\alpha,\beta,\sigma,\sigma'}[g_1\psi_{\alpha\sigma}^{\dagger}\psi_{\beta\sigma'}^{\dagger}\psi_{\alpha\sigma'}\psi_{\beta\sigma}+g_2\psi_{\beta\sigma}^{\dagger}\psi_{\alpha\sigma'}^{\dagger}\psi_{\alpha\sigma'}\psi_{\beta\sigma}\nonumber\\
&+&g_3\psi_{\alpha\sigma}^{\dagger}\psi_{\alpha\sigma'}^{\dagger}\psi_{\beta\sigma'}\psi_{\beta\sigma}+g_5\psi_{\beta\sigma}^{\dagger}\psi_{\beta'\sigma'}^{\dagger}\psi_{\alpha'\sigma'}\psi_{\alpha\sigma}\nonumber\\
&+&\tilde g_5\psi_{\beta\sigma}^{\dagger}\psi_{\alpha'\sigma'}^{\dagger}\psi_{\beta'\sigma'}\psi_{\alpha\sigma}]-\frac{1}{2}\sum_{\alpha,\sigma,\sigma'}[\tilde g_1\psi_{\alpha'\sigma}^{\dagger}\psi_{\alpha\sigma'}^{\dagger}\psi_{\alpha'\sigma'}\psi_{\alpha\sigma}\nonumber\\
&+&\tilde g_2\psi_{\alpha\sigma}^{\dagger}\psi_{\alpha'\sigma'}^{\dagger}\psi_{\alpha'\sigma'}\psi_{\alpha\sigma}+g_4\psi_{\alpha\sigma}^{\dagger}\psi_{\alpha\sigma'}^{\dagger}\psi_{\alpha\sigma'}\psi_{\alpha\sigma}],\label{L}
\end{eqnarray}
where $\alpha$ and $\beta$ are patch indices, $\alpha'$ labels the patch conjugate to $\alpha$, $\sigma=\uparrow,\downarrow$ is the spin index. The sum over $\alpha$ and $\beta$ is taken over only non-conjugate patches. Note that the Umklapp scattering, $g_3$, is forbidden because it does not conserve momentum modulo a reciprocal lattice vector \cite{Shtyk17}.

The building blocks of the RG analysis are the polarisation operators in the particle-particle and particle-hole channels, shown in the two top diagrams of Fig. \ref{Fig3} (b) respectively, at zero  momentum transfer  and at momentum transfer $\mathbf Q_{\alpha \beta}$ between two patches $\alpha$ and $\beta$. They can be calculated as
\begin{equation}
\Pi_{pp}(\mathbf q)=T\int_pG(i\omega_n,\mathbf p+\mathbf q)G(-i\omega_n,-\mathbf p),\label{Pipp}
\end{equation}
\begin{equation}
\Pi_{ph}(\mathbf q)=-T\int_pG(i\omega_n,\mathbf p+\mathbf q)G(i\omega_n,\mathbf p),\label{Piph}
\end{equation}
where $\int_p...=\sum_n\int d^2p$, $\omega_n=\pi (2n+1)T$, and $G(\mathbf p,i\omega_n)=[i\omega_n-\epsilon(\mathbf p)+\mu]^{-1}$ is the fermionic Matsubara Green's function. 
For energies (\ref{e1}), the polarisation operators can be evaluated as:
\begin{equation}
\Pi_{ph}(\mathbf Q_{\alpha,\alpha'})=\Pi_{ph}(0)=\nu_0\ln \left (\frac{\Lambda}{max\{T,|\mu|\}} \right ),\label{ph0}
\end{equation} 
\begin{equation}
\Pi_{ph}(\mathbf Q_{\alpha,\beta})=A\nu_0\ln \left (\frac{\Lambda}{max\{T,|\mu|\}} \right ),\label{phq}
\end{equation} 
\begin{equation}
\Pi_{pp}(\mathbf Q_{\alpha,\alpha'})=\Pi_{pp}(0)=\frac{\nu_0}{2}\ln \left (\frac{\Lambda}{max\{T,|\mu|\}} \right )\ln \left (\frac{\Lambda}{T} \right ),\label{pp0}
\end{equation} 
\begin{equation}
\Pi_{pp}(\mathbf Q_{\alpha,\beta})=B\nu_0\ln \left (\frac{\Lambda}{max\{T,|\mu|\}} \right ),\label{ppq}
\end{equation} 
where $A=\frac{\sqrt 3}{5}\ln\left (\frac{37+20 \sqrt 3}{13}\right )\approx0.59$ and $B=\frac{6}{\sqrt 39}\left [\pi-\arctan\left (2\sqrt{\frac{3}{13}}\right)\right]\approx2.28$. Note that the polarisation operators at the momentum transfer connecting two conjugate saddle points are equal to the ones at zero momentum transfer due to energy degeneracy. Similarly to the case of monolayer graphene doped to the saddle (M-) points, $\Pi_{pp}(0)\propto \ln[\Lambda/T]^2$ at $\mu\ll T$ \cite{Nandkishore12,Gonzalez08}, however, in the present case we find that $\Pi_{ph}(\mathbf Q_{\alpha,\beta})$ is linear in $\ln(\Lambda/T)$  contrary to the monolayer graphene, where the dependence can be quadratic. This difference is because, in the case of twisted bilayer graphene there is no nesting in Fermi surfaces, again in contrast with monolayer graphene. 
The relevant Feynman diagrams are shown in Fig. \ref{Fig3} (a) and (b). We then obtain the following RG equations in one-loop approximation
\begin{eqnarray}
\frac{dg_1}{dy}&=&2d_0g_1(g_1+\tilde g_1+g_4)+2d_1[g_1(g_2-g_1)+g_5(\tilde g_5-g_5)]\nonumber\\
&-&2d_3g_1g_2,\label{g1}\\
\frac{d\tilde g_1}{dy}&=&d_0[4g_1^2+2\tilde g_1(\tilde g_2+g_4-\tilde g_1)]\nonumber\\
&-&2d_4(y)(\tilde g_1\tilde g_2+g_5\tilde g_5), \label{g1p}\\
\frac{dg_2}{dy}&=&2d_0[g_2(\tilde g_1-2\tilde g_2-2g_2-g_4)+g_1(\tilde g_2+g_4)+2g_1g_2]\nonumber\\
&+&2d_1(g_2^2+\tilde g_5^2)-2d_3(g_1^2+g_2^2),\label{g2}\\
\frac{d\tilde g_2}{dy}&=&d_0[\tilde g_2(\tilde g_2-2g_4)+8g_2(g_1-g_2)+2\tilde g_1g_4]\nonumber\\
&-&d_4(y)(\tilde g_1^2+\tilde g_2^2+2g_5^2+2\tilde g_5^2), \label{g2p}\\
\frac{dg_4}{dy}&=&d_0[g_4^2-8g_2^2-2\tilde g_2^2+4g_1(g_1+2g_2)+\tilde g_1(\tilde g_1+2\tilde g_2)]\nonumber\\
&-&d_4(y)g_4^2,\label{g4}\\
\frac{dg_5}{dy}&=&2d_1[2g_2g_5+g_1(\tilde g_5-2g_5)]\nonumber\\
&-&d_4(y)(2\tilde g_1\tilde g_5+2\tilde g_2g_5+g_5^2+\tilde g_5^2),\label{g5}\\
\frac{d\tilde g_5}{dy}&=&2d_1[2g_2\tilde g_5+g_1(g_5-2\tilde g_5)]\nonumber\\
&-&d_4(y)(2\tilde g_1 g_5+2\tilde g_2\tilde g_5+g_5^2+\tilde g_5^2),\label{g5p}
\end{eqnarray}
where $y=\Pi_{ph}(0)/\nu_0$ and $g_i\rightarrow g_i\nu_0$ is dimensionless. We also define $d_0=1$,  $d_1=\frac{d\Pi_{ph}(\mathbf Q_{\alpha\beta})}{d\Pi_{ph}(0)}=A$, $d_3=\frac{d\Pi_{pp}(\mathbf Q_{\alpha\beta})}{d\Pi_{ph}(0)}=B$, and $d_4(y)=\frac{d\Pi_{pp}(0)}{d\Pi_{ph}(0)}=\ln (\Lambda/T)/2\equiv x/2$. Given that, for $T\gg |\mu|$, $x=y$, and, for $T\ll |\mu|$, $x\gg y=\ln(\Lambda/|\mu|)$, we interpolate $d_4(y)$ as
\begin{equation}
 d_4(y)=\frac{yz}{2(z-y)}, \label{d4}
 \end{equation}
 where $z=\ln(\Lambda/|\mu|)$.

\begin{figure}[h]
\includegraphics[width=0.45\textwidth]{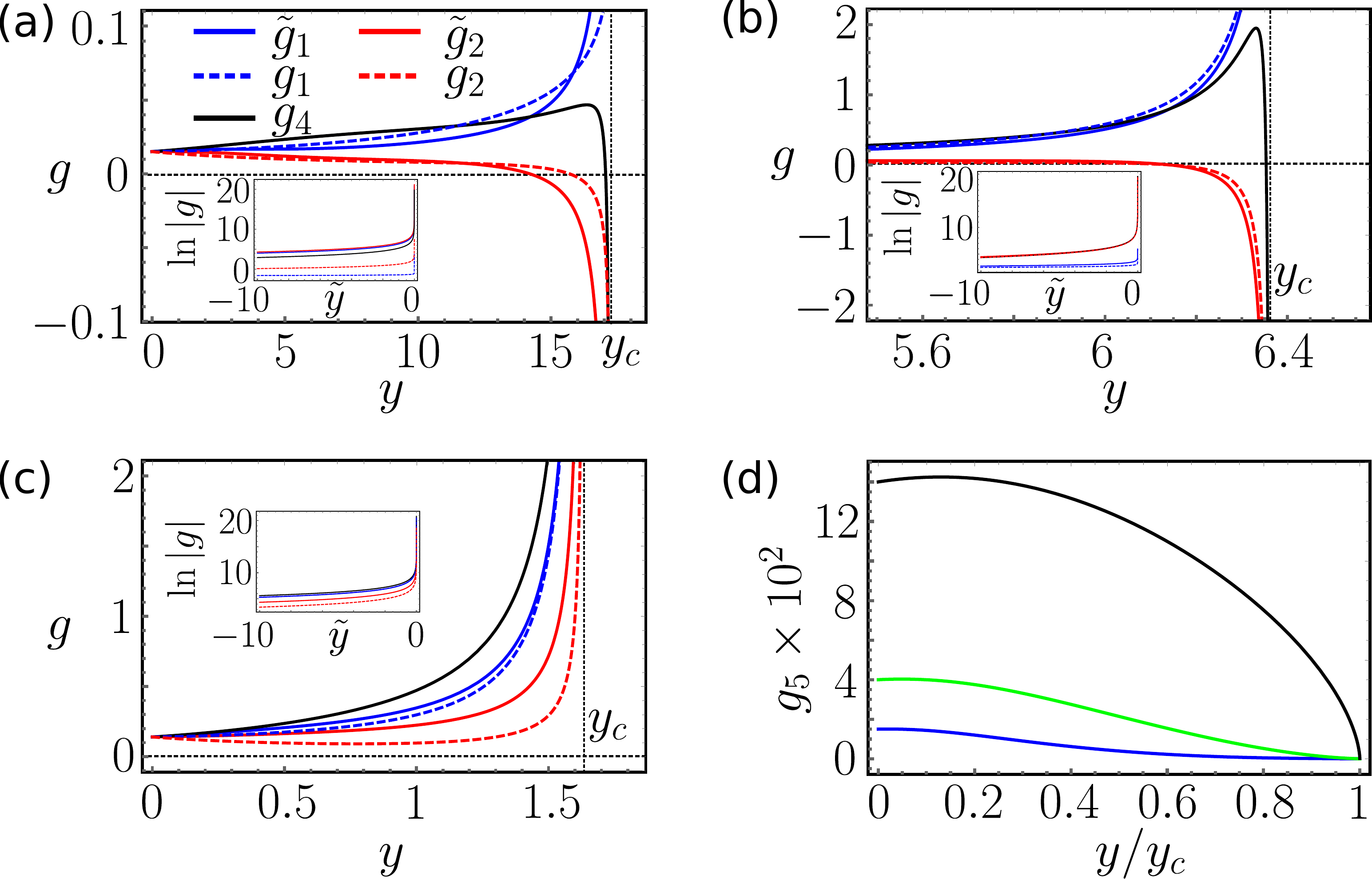}
\caption{(a)-(c) Flow of the coupling constants  with renormalisation group scale $y$ starting from repulsive interaction $g_i(0)=g_0>0$ for (a) $g_0=0.015$; (b) $g_0=0.04$, and (c) $g_0=0.14$. The chemical potential  $\mu=0$ corresponds to the van Hove filling. Insets: $\ln|g_i|$ as a function of $\tilde y=10^4(y-y_c)$ in the vicinity of $y_c$, demonstrating $|g_1|,|g_2|\ll |g_4|,|\tilde g_1|,|\tilde g_2|$ in case (a) and $|g_1|,|\tilde g_1|\ll |g_2|,|\tilde g_2|, |g_4|$ in case (b). (d) $g_5=\tilde g_5$ as a function of $y/y_c$ for $g_0=0.015$ (blue),  $g_0=0.04$ (green), and  $g_0=0.14$ (black).}
\label{Fig4}\end{figure} 

An alternative way to get the RG equations, with a set of different assumptions, is presented in Appendix A. This procedure retains the structure of the RG equations with a different definition of the coefficients and leads to the same physical results. For completeness, all the results are included in Appendix A.

\section{Results of the RG analysis.} 

\begin{figure}[h]
\includegraphics[width=0.45\textwidth]{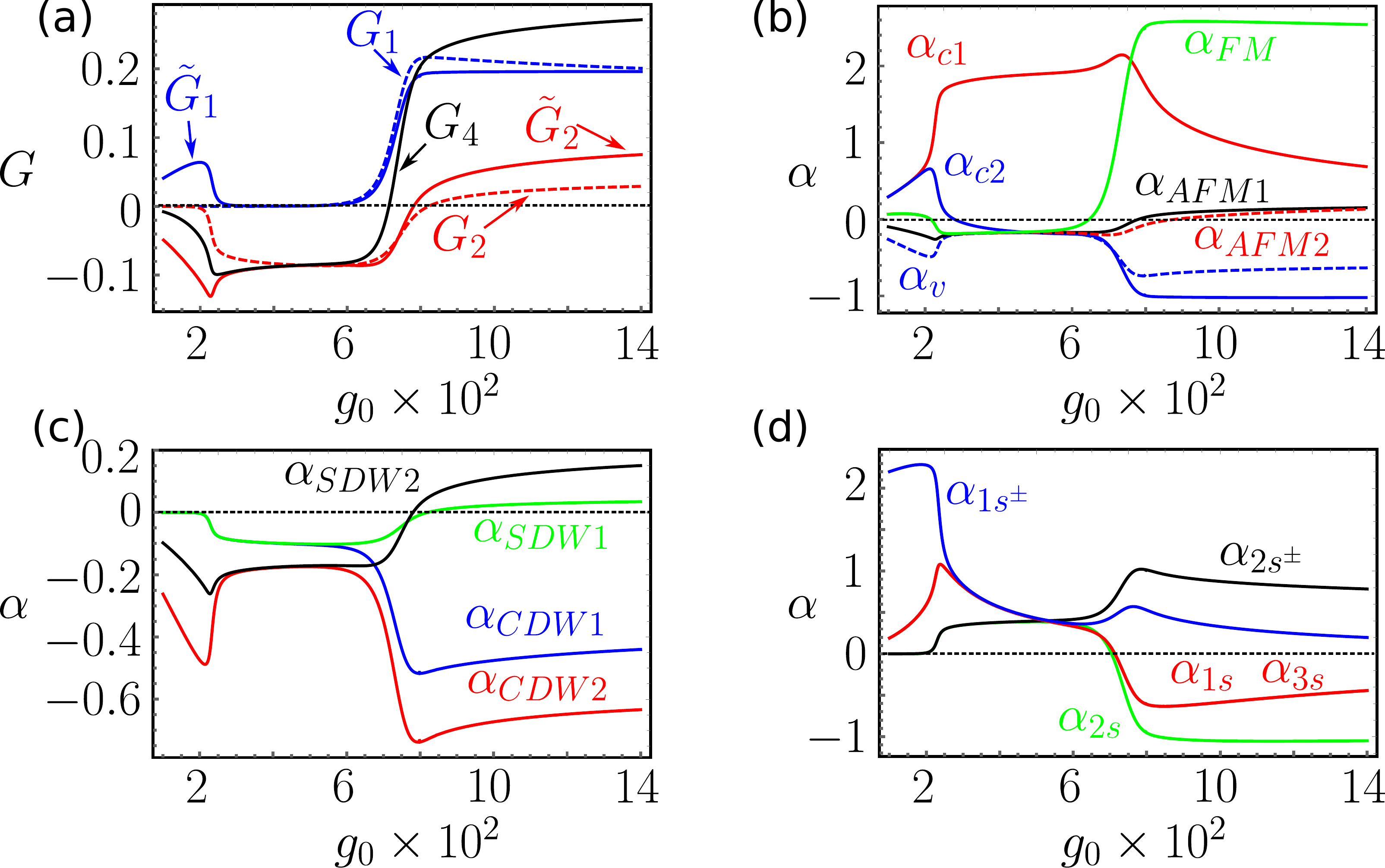}
\caption{Parameters of susceptibilities at the van Hove filling for  $g_i(0)=g_0>0$. (a) $G_i$ as  functions of $g_0$. (b)-(d) $\alpha_i$ as functions of $g_0$ for (b) uniform densities; (c) spin and charge density waves; and (d) superconductivity.   }
\label{Fig5}\end{figure}

In Fig. \ref{Fig4}, we show the numerical solutions of Eqs. (\ref{g1}) - (\ref{g5p}) for $g_i(0)=g_0>0$ and $\mu=0$. We  found that $g_5$ and $\tilde g_5$ are irrelevant (see Fig. \ref{Fig4} (d)). The remaining couplings diverge at a scale $y_c\propto g_0^{-1/2}$. For $g_0<g_0^{(1)}\approx 0.024$, the main contribution comes from the scattering between conjugate points described by $\tilde g_1$ and $\tilde g_2$, as well as intra-patch interaction $g_4$, with $\tilde g_1$ flowing to repulsion and $\tilde g_2$ and $g_4$ flowing to attraction, as shown in Fig. \ref{Fig4} (a). For $g_0^{(1)}<g_0<g_0^{(2)}\approx 0.075$,  the main contribution arises from $\tilde g_2$, $g_2$, and $g_4$, which flow to attraction, as shown in Fig. \ref{Fig4} (b). At $g_0>g_0^{(2)}$ all contributions are of the same order, however, the parameters $g_i$ flow to repulsion (see Fig. \ref{Fig4} (c)). To study the couplings for the whole range of $y$, we note that close to $y_c$, the relevant coupling can be cast as $g_i=G_i/(y_c-y)$. Substituting it into Eqs. (\ref{g1}) - (\ref{g4}), we find that $G_i$ satisfies simultaneous polynomial equations, which can be solved numerically to confirm our observations, as we show in Fig. \ref{Fig5} (a). 

The nature of instabilities can be identified with the help of the relevant susceptibilities \cite{Furukawa98,Nandkishore12,Shtyk17} as the most divergent susceptibility corresponds to the leading instability. Presenting the susceptibilities, $\chi_i$, close to  $y_c$, as $\chi_i=(y_c-y)^{-\alpha_i}$, the leading instability can be found as the one with maximal positive $\alpha_i$.

We then introduce infinitesimal test vertices and study their renormalisation, described by three-leg diagrams in Fig. \ref{Fig3} (c).  The test vertex for the instabilities due to uniform densities, $\delta \mathcal L=\sum_{\sigma\alpha}n_{\sigma,\alpha}\psi_{\sigma\alpha}^{\dagger}\psi_{\sigma\alpha}$, can be renormalised in one-loop approximation according to
\begin{eqnarray}
\frac{dn_{\sigma\alpha}}{dy}&=&d_0\{-g_4n_{\bar{\sigma}\alpha}+(\tilde g_1-\tilde g_2)n_{\sigma\alpha'}-\tilde g_2n_{\bar{\sigma}\alpha'}\nonumber\\
&+&\sum_{\beta\neq\alpha,\alpha'}[(g_1-g_2)n_{\sigma\beta}-g_2n_{\bar{\sigma}\beta}]\},\label{unif}
\end{eqnarray}
where $\bar{\sigma}=-\sigma$, and primes again mark conjugate saddle points. The right-hand side of Eq. (\ref{unif}) can be casted in a $12\times 12$ matrix in the basis $\{n_{\uparrow1},n_{\downarrow1},n_{\uparrow1'},n_{\downarrow1'}...\}$, whose eigenvalues, $\gamma_i$, are related to the susceptibilities as $\alpha_i=2\gamma_i$. We find six distinct eigenvalues corresponding to charge, $\alpha_{c1}$ and $\alpha_{c2}$, valley, $\alpha_v$, antiferromagnetic, $\alpha_{AFM1}$ and $\alpha_{AFM2}$, and ferromagnetic, $\alpha_{FM}$, instabilities:
\begin{eqnarray}
\alpha_{c1}&=&2d_0(\tilde G_1-2\tilde G_2-G_4+4G_1-8G_2)\label{c1},\\
\alpha_{c2}&=&2d_0(\tilde G_1-2\tilde G_2-G_4-2G_1+4G_2)\label{c2},\\
\alpha_{v}&=&2d_0(-\tilde G_1+2\tilde G_2-G_4)\label{v},\\
\alpha_{AFM1}&=&2d_0(-\tilde G_1+G_4)\label{AFM1},\\
\alpha_{AFM2}&=&2d_0(\tilde G_1+G_4-2G_1)\label{AFM2},\\
\alpha_{FM}&=&2d_0(\tilde G_1+G_4+4G_1)\label{FM}.
\end{eqnarray} 
Next we turn to spin and charge density wave instabilities, for which the test vertex, $\delta \mathcal L=\sum_{\sigma\mathbf Q}n_{\sigma\mathbf Q}\psi_{\sigma\beta}^{\dagger}\psi_{\sigma\alpha}+H.C.$, where we use $\mathbf Q\equiv \mathbf Q_{\alpha\beta}$. The renormalisation of $n_{\sigma\mathbf Q}$ can be obtained from the one-loop equations for non-conjugate patches,
\begin{eqnarray}
\frac{dn_{\sigma\mathbf Q}}{dy}=d_1[(g_2-g_1)n_{\sigma\mathbf Q}-g_1n_{\sigma\mathbf Q}]\label{nq},
\end{eqnarray} 
while for the conjugate ones
\begin{eqnarray}
\frac{dn_{\sigma\mathbf Q'}}{dy}=d_0[(\tilde g_2-\tilde g_1)n_{\sigma\mathbf Q'}-\tilde g_1n_{\sigma\mathbf Q'}]\label{nqp},
\end{eqnarray} 
where in our notation $\mathbf Q'\equiv \mathbf Q_{\alpha\alpha'}$.
These equations yield charge density wave (CDW) and spin density wave (SDW) instabilities as
\begin{eqnarray}
\alpha_{CDW1}&=&2d_1(-2G_1+G_2),\label{CDW1}\\
\alpha_{CDW2}&=&2d_0(-2\tilde G_1+\tilde G_2),\label{CDW2}\\
\alpha_{SDW1}&=&2d_1G_2,\label{SDW1}\\
\alpha_{SDW2}&=&2d_0\tilde G_2,\label{SDW2}
\end{eqnarray}
where index $1$ ($2$) corresponds to the density waves developed on non-conjugate (conjugate) patches.

\begin{figure}[h]
\includegraphics[width=0.45\textwidth]{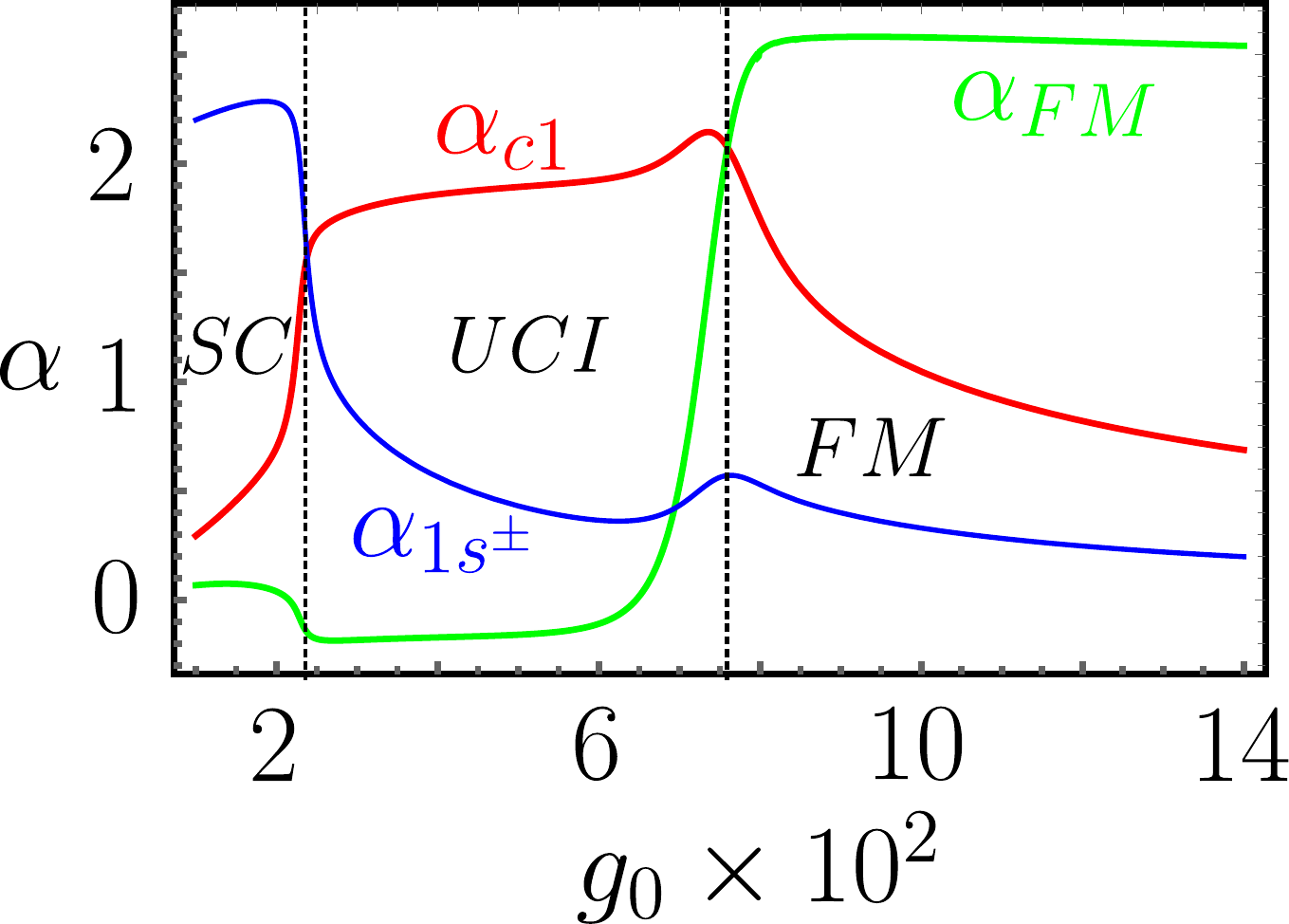}
\caption{Phase diagram of the system at van Hove filling: at $g_0< g_0^{(1)}\approx 0.024$, the leading instability is $s^{\pm}$ superconductivity (SC) with the order parameter changing sign along the path connecting two conjugate patches. At $g_0^{(1)}< g_0< g_0^{(2)}\approx 0.075$, the most rapidly divergent is uniform charge susceptibility leading to uniform charge instability phase (UCI), and at $g_0>g_0^{(2)}$ the leading instability is due to ferromagnetism (FM). }
\label{Fig6}\end{figure}

To study superconductivity we introduce intra- and inter-patch vertices, $\delta\mathcal L_{intra}=\sum_{\alpha}\Delta_{\alpha}\psi_{\uparrow\alpha}\psi_{\downarrow\alpha}+H.C.$, and $\delta\mathcal L_{inter}=\sum_{\mathbf Q}(\Delta_{\mathbf Q}^{(1)}\psi_{\uparrow\alpha}\psi_{\downarrow\beta}+\Delta_{\mathbf Q}^{(2)}\psi_{\uparrow\beta}\psi_{\downarrow\alpha})+h.c.$, which renormalise according to the following equation:
\begin{equation}
\frac{d\Delta_{\alpha}}{dy}=-d_4G_4\Delta_{\alpha},\label{sc1}
\end{equation} 
and ($j\neq i$)
\begin{equation}
\frac{d\Delta_{\mathbf Q}^{(i)}}{dy}=-d_3(G_2\Delta_{\mathbf Q}^{(i)}+G_1\Delta_{\mathbf Q}^{(j)}),\label{sc2}
\end{equation} 
\begin{equation}
\frac{d\Delta_{\mathbf Q'}^{(i)}}{dy}=-d_4(\tilde G_2\Delta_{\mathbf Q'}^{(i)}+\tilde G_1\Delta_{\mathbf Q'}^{(j)}).\label{sc3}
\end{equation} 
The prime denotes conjugate patches. Note that, in contrast with \cite{Nandkishore12}, the equation for intra-patch order parameter is diagonal. This is because the Umklapp  processes $g_3$ are forbidden. This excludes all symmetries of the superconductive order parameters,  except from $s$ and $s^{\pm}$. This is a major difference with respect to results on single layer graphene.
The intra-patch order parameter has $s-$ wave symmetry and is given by
\begin{equation}
\alpha_{1s}=-2d_4G_4.\label{s1}
\end{equation}
However, for the inter-patch order parameter we find both $s-$ wave and $s^{\pm}-$ wave symmetry. The former is given by
\begin{equation}
\alpha_{2s}=-2d_3(G_1+G_2),\label{2s}
\end{equation}
for non-conjugate patches, and
\begin{equation}
\alpha_{3s}=-2d_4(\tilde G_1+\tilde G_2),\label{3s}
\end{equation}
for conjugate patches.
For the $s^{\pm}$ order parameter we find
\begin{equation}
\alpha_{1s^{\pm}}=2d_4(\tilde G_1-\tilde G_2),\label{1spm}
\end{equation}
for conjugate patches, and 
\begin{equation}
\alpha_{2s^{\pm}}=2d_3(G_1- G_2),\label{2spm}
\end{equation}
for non-conjugate patches.
In Figs. \ref{Fig5} and \ref{Fig6} we compare $\alpha_i$ for all potential instabilities at $\mu=0$ and for $g_i(0)=g_0>0$.  The leading instability at $g_0<g_0^{(1)}$ is inter-patch $s^{\pm}$ superconductivity corresponding to the coupling between conjugate patches with the order parameter changing sign along the path connecting two conjugate patches (see Fig. \ref{Fig6}). For $g_0^{(1)}< g_0< g_0^{(2)}$, the most divergent is the uniform charge susceptibility, which corresponds to the UCI phase, often referred to as phase separation (PS) between two states with different electronic densities as has been observed around transitions to a Mott insulating state \cite{Kotliar02,Aichhorn07,Macridin06,Eckstein07,Misawa14,Otsuki14} (see Fig. \ref{Fig6}).  And, finally, for $g_0>g_0^{(2)}$ the leading instability is due to ferromagnetism. The dependence of $\alpha_i$ on the initial condition $g_0$ might seam unusual, however, as we show in  Appendix A, this can be explained by the dependence of the parameters $d_i$ on $y_c$ and consequently, on $g_0$.

\begin{figure}[h]
\includegraphics[width=0.45\textwidth]{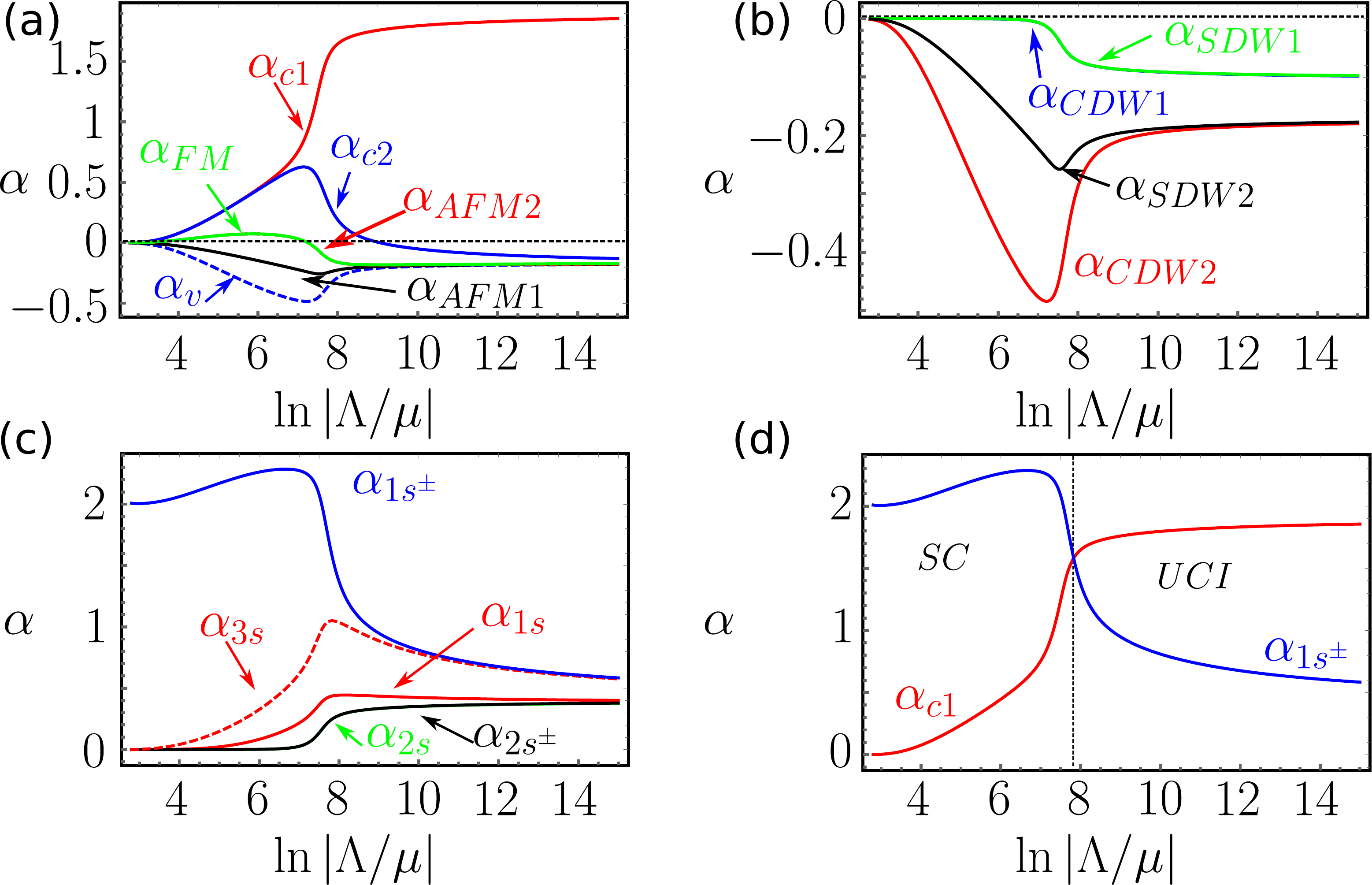}
\caption{Parameters of susceptibilities close to van Hove filling for $g_0=0.055$. (a)-(c) $\alpha_i$ as functions of $z=\ln |\Lambda/\mu|$. (d) Leading instabilities as functions of $z$ showing the two dominating phases:   $s^{\pm}$superconductivity (SC) with the order parameter changing sign along the path connecting the two conjugate patches and the uniform charge instability phase (UCI) at $z_c\approx 7.5$.   }
\label{Fig7}\end{figure} 
For finite chemical potential, \textit{i.e.} away from the van Hove point, the RG equations (\ref{g1})-(\ref{g5p}), as well the susceptibilities (\ref{c1})-(\ref{2spm}) do not change, however, we use the approximation (\ref{d4}) for the parameter $d_4$. This allows us to compare the susceptibilities for different order parameters as a function of $z=\ln(\Lambda/|\mu|)$. In Fig. \ref{Fig7}, we show $\alpha_i$ for $g_0=0.055$, corresponding to the UCI phase under the van Hove doping. We found that for $z<z_c\approx 7.5$, the leading instability is the inter-patch $s^{\pm}$ superconductivity corresponding to the coupling between conjugate patches with the order parameter changing sign along the path connecting two conjugate patches, followed by the UCI state at $z>z_c$.
To estimate the transition temperature, we use the same approximation as for ({\ref{d4}) and cast $y_c$ as $y_c=\frac{x_cz}{z-x_c}$, where $x_c=\ln(\Lambda/T_c)$. We then evaluate $y_c$ as a function of $z$ and extract $T_c$. The chemical potential can be calculated in terms of doping electronic density, $n$, as  \cite{CaoMott18} $\mu=\hbar v^*\sqrt{n\pi/2}-\epsilon_0$, for which the van Hove doping is determined by the electronic density $n_0=\frac{m^2v^{*2}}{2\pi\hbar^2}$. This allows us to present our result as the phase diagram in Fig. \ref{Fig8}. Note that the symmetry of the phase diagram is due to the fact that $\mu$ changes sign at $n=n_0$, however, it is $|\mu|$ but not $\mu$ that enters all the expression for the polarisation operators. Note that the width of the UCI phase is determined by the value of the coupling constant $g_0$, in accordance with Fig. \ref{Fig6}. For $g_0<g_0^{(1)}$, the UCI phase  is absent, and the system becomes superconducting for all doping charge densities close to $n_0$. At $g_0=g_0^{(1)}$, the UCI phase appears at $n=n_0$ and expands  as $g_0$ grows. For $g_0\gtrsim g_0^{(2)}$, the ferromagnetic phase appears at $n\simeq n_0$.

\begin{figure}[h]
\includegraphics[width=0.45\textwidth]{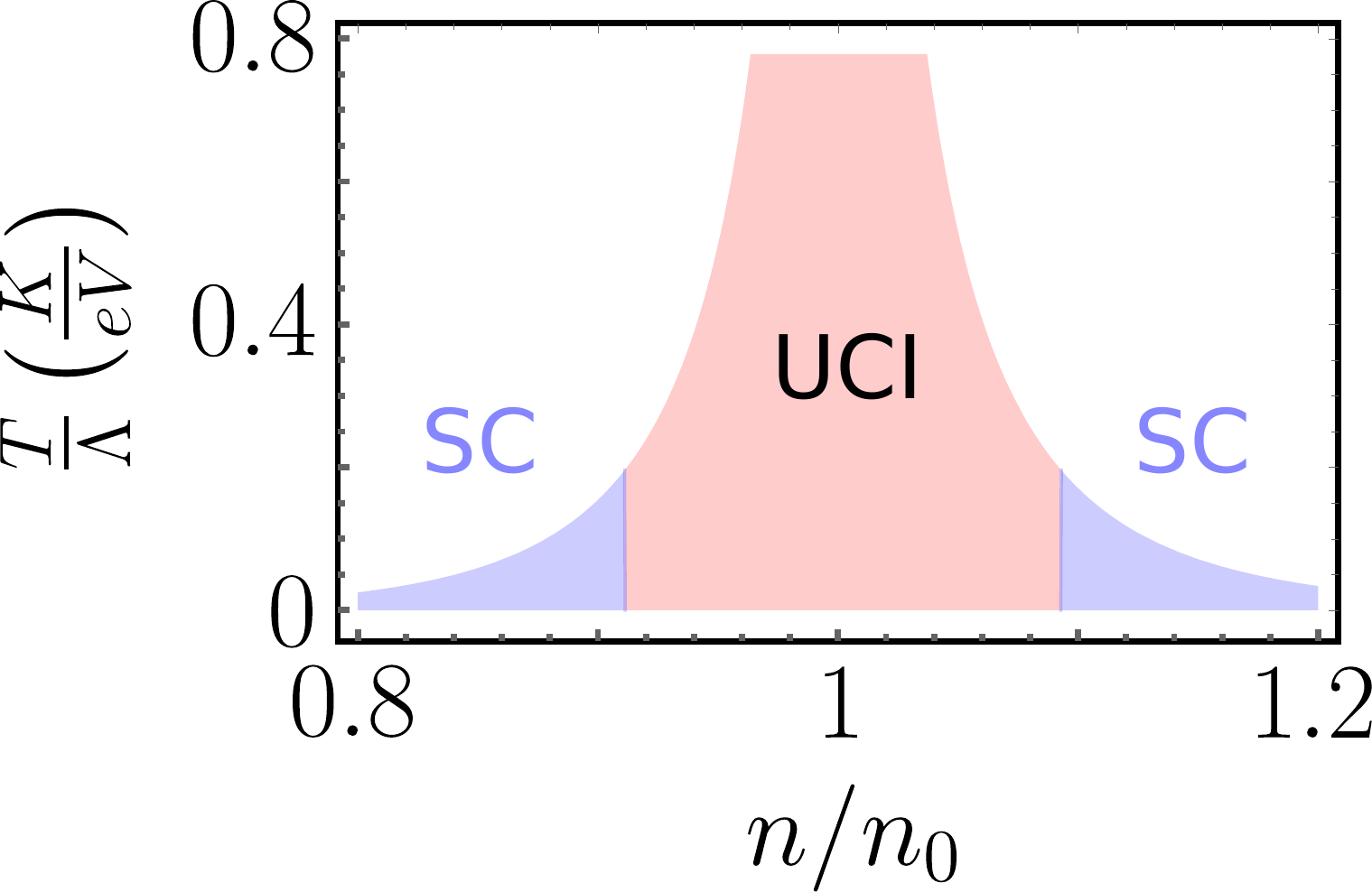}
\caption{Phase diagram for the system close to the van Hove filling as a function of doping charge density, $n$, relative to the doping charge density, $n_0$, corresponding to the van Hove filling for the set of parameters: $g_0=0.055$, and  $\epsilon_0=mv^{*2}/2=8\times10^{-3}\Lambda$. The two   $s^{\pm}$ superconducting phases,  with the order parameter changing sign along the path connecting two conjugate patches, are separated by a phase with a uniform charge instability. }
\label{Fig8}\end{figure} 


\section{Discussion.} 

The physics and the corresponding phase diagram in Fig. \ref{Fig8} is relevant to the experimental results of \cite{CaoSC18} obtained for $\nu=1$ per moire cell. In the region near $n_0$, our results suggest that there is  a phase of a uniform charge instability, identified as the state of phase separation. In the context of Hubbard model, this state of phase separation is seen in a number of reported calculations using a variety of techniques  \cite{Kotliar02,Aichhorn07,Macridin06,Eckstein07,Misawa14,Otsuki14}. As a result of the coexistence of metallic and insulating regions, the transport in this particular phase is through percolation. Therefore it is possible that in the range of values of densities and temperature  it can display insulating behavior. This phase must be investigated further. The results have been verified by an alternative way of using the RG procedure as summarised in Appendix A.

In the case when the inter-valley coupling, $g_0^{inter}$, is negligible in comparison with the intra-valley one, $g_0^{intra}$,  the system can be modelled by taking into account only the fermions living in three patches around van Hove singularities that belong to the same valley. In this case the two-fermion scattering is described by three distinct interactions in the low-energy theory, $g_1$, $g_2$, and $g_4$. The energy dispersions in the patches are all distinct and given by Eqs. (\ref{e1}). In this case, as we show in Appendix B, the phase diagram is similar to the one in Fig. \ref{Fig8}: the UCI phase is sandwiched between the two superconducting phases (see Fig. \ref{Fig2A} in Appendix B).   However, the symmetry of the order parameter of the SC phase is now different. We predict, that in the case $g_0^{inter}\ll g_0^{intra}$, it is of $s^{++}-$ symmetry, in contrast with $s^{\pm}$ order parameter predicted for $g_0^{inter}=g_0^{intra}$, \textit{i.e}, it does not change sign along the path connecting two patches. Note also that in our calculations we considered only nearest neighbor interactions corresponding to particle-hole symmetry. Taking into account further neighbors in interaction will break the particle hole symmetry, however, it will not change the results qualitatively.

In conclusion, we have investigated a microscopic model that takes into account interactions among electrons that live around the points of van Hove singularities formed in twisted bilayer graphene near filling factor $\nu=1$ and found the different phases as a function of electron density. We found superconductivity which displays a reentrant behavior as a function of electronic density. We predict that the order parameter symmetry is $s^{\pm}$, while if the inter-valley coupling is negligible it is of $s^{++}$ symmetry. The phase in the middle of the two superconducting phases is characterised by a divergence of charge susceptibility at $q=0$. This signals a phase of coexistence of metallic and insulating regions in a phase separation. The other phase that appears in the phase diagram is the ferromagnetic one. It is worth mentioning that the width of the two superconducting phases can be tuned according to the values of $g_0$ therefore extending the regions. As a result, the physics associated with different filling factors is quite different, as different singularities in DOS and different nesting conditions exist. Our theory can accommodate recent experiments \cite{Dean} where pressure has been used to tune the system, which is work in progress.

\section*{Acknowledgements}
We are grateful to Andrey Chubukov for his questions and comments, that led us to verify the results using an alternative RG procedure. We also thank Dima Efremov, Mark Greenaway and Andreas Rost for useful discussions on related problems. This work was supported by the EPSRC through the grant EP/P002811/1.

\appendix
\section{Alternative RG equations}
Alternatively, we can obtain the RG equations by differentiating $g_i$ with respect to the renormalization group "time", $y=2\Pi_{pp}(0)/\nu_0$, which, for $|\mu|\ll T$, is quadratic in $\ln|\Lambda/T|$. This procedure leads again to the same equations  Eqs. (\ref{g1})-(\ref{g5p}) with different parameters $d_i$: $d_0=y^{-1/2}$,  $d_1= A y^{-1/2}$, $d_3=By^{-1/2}$, and $d_4=1$. We solve Eqs. (\ref{g1})-(\ref{g5p}) with the new parameters $d_i$ following the method of \cite{Nandkishore12}; because the renormalization group equations flow to
strong coupling at a finite scale $y_c$, we treat  $d_0(y_c)$ as a parameter in our calculations. However, in contrast with \cite{Nandkishore12}, we keep the sub-leading terms, $d_1$ and $d_3$, and assume that  $d_1/d_0=A=const$ and $d_3/d_0=B=const$.  We then calculate $g_i=\frac{G_i}{y_c-y}$ and find the  susceptibilities as a function of $d_0(y_c)$. The results are shown in Fig. \ref{Fig1B}.  The phase diagram that is extracted from the results of this method Fig. \ref{Fig1B} (d), has the same structure as the one presented in Fig. \ref{Fig6}:  a UCI phase is sandwiched  between the $s^{\pm}$ superconducting phase and the FM-phase. We also  find that $y_c\propto g_0^{-1}$, which explains the dependence of $\alpha_i$ on $g_0$ in Fig. \ref{Fig6}: the solutions of Eqs. (\ref{g1}) -(\ref{g5p}) with constant $d_i$ are independent of $g_0$, however,  $d_0$ does depend on $g_0$ as $d_0\propto \sqrt {g_0}$, which, in turn, carries the dependence on the initial condition to $g_i$ and $\alpha_i$.  
\begin{figure}[h]
\includegraphics[width=0.45\textwidth]{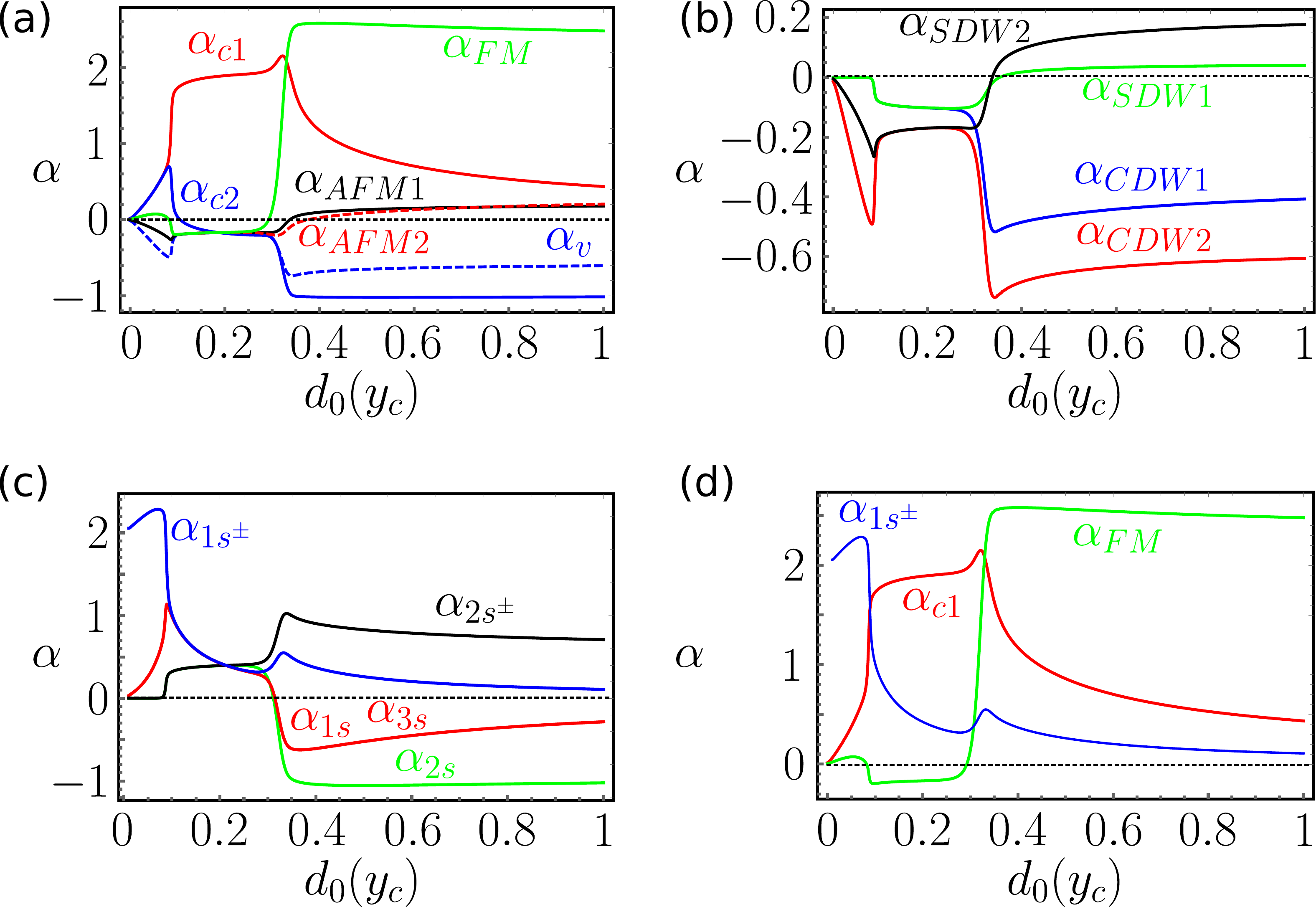}
\caption{(a) - (c) Parameters of susceptibilities at the van Hove filling as functions of $d_0(y_c)$ for $g_i(0)=g_0>0$ and equal inter- and intra-valley scatterings  $g_0^{inter}= g_0^{intra}$. (d) Phase diagram with leading instabilities corresponding to $s^{\pm}-$wave superconductivity, UCI phase, and  FM phase.  }
\label{Fig1B}\end{figure} 

\section{The case of negligible inter-valley scattering}
Here we consider the limit of the screened Coulomb interaction, $U(\mathbf k)$, being short-ranged in the $\mathbf k-$ space, \textit{i.e.} $U(\Delta \mathbf K_s)\ll U(0)$. In this limit, the inter-valley scattering is negligible, and the low-energy Lagrangian can be expressed in terms of the three distinct interaction, $g_1$, $g_2$, and $g_4$, coupling only the patches belonging to the same valley:
\begin{eqnarray}
\mathcal{L}&=&\sum_{\alpha,\sigma}\psi_{\alpha\sigma}^{\dagger}(\partial_{\tau}-\epsilon_k+\mu)\psi_{\alpha\sigma}\nonumber\\
&-&\frac{1}{2}\sum_{\alpha,\beta,\sigma,\sigma'}[g_1\psi_{\alpha\sigma}^{\dagger}\psi_{\beta\sigma'}^{\dagger}\psi_{\alpha\sigma'}\psi_{\beta\sigma}+g_2\psi_{\beta\sigma}^{\dagger}\psi_{\alpha\sigma'}^{\dagger}\psi_{\alpha\sigma'}\psi_{\beta\sigma}]\nonumber\\
&-&\frac{1}{2}\sum_{\alpha,\sigma,\sigma'}g_4\psi_{\alpha\sigma}^{\dagger}\psi_{\alpha\sigma'}^{\dagger}\psi_{\alpha\sigma'}\psi_{\alpha\sigma},\label{LA}
\end{eqnarray}
where patches $\alpha$ and $\beta$ belong to the same valley. 

\begin{figure}[h]
\includegraphics[width=0.45\textwidth]{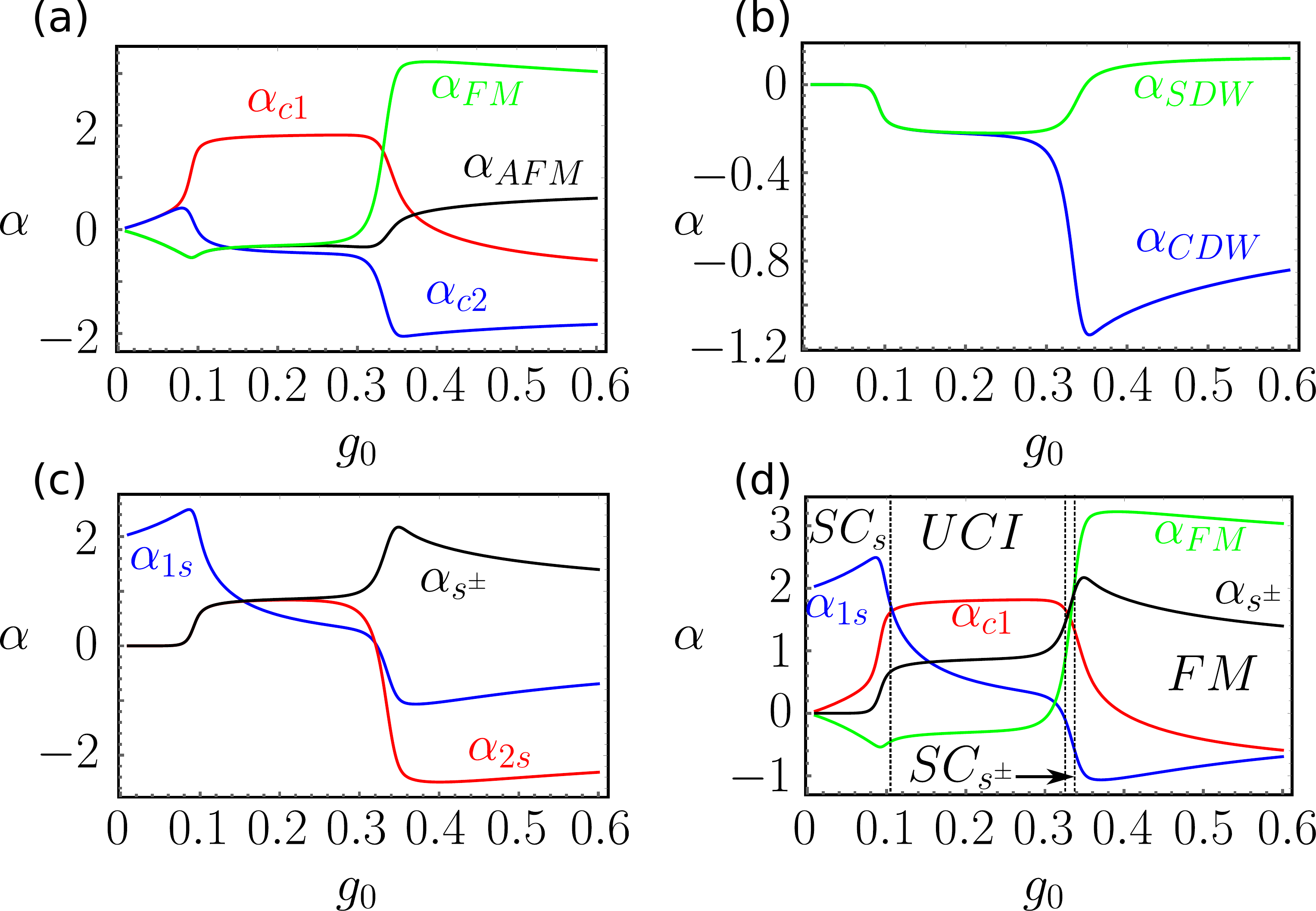}
\caption{(a) - (c) Parameters of susceptibilities at van Hove filling as functions of $g_i(0)=g_0>0$ for $g_0^{inter}\ll g_0^{intra}$. (d) Phase diagram with leading instabilities corresponding to $s-$wave superconductivity ($SC_s$), UCI phase, $s^{\pm}-$superconductivity ($SC_{s^{\pm}}$), and FM phase.  }
\label{Fig1A}\end{figure} 
We compare the susceptibilities for the case of $\mu=0$ in Fig. \ref{Fig1A}. We found that for $g_0\lesssim 0.1$, the leading instability is intra-patch $s-$wave superconductivity, followed by UCI phase for $0.1\lesssim g_0\lesssim 0.32$, $s^{\pm}-$ wave superconductivity at the narrow range  $0.32\lesssim g_0\lesssim 0.34$, and, finally, FM phase for $g_0\gtrsim 0.34$, as we show in Fig. \ref{Fig1A} (d).

Using the Feynman diagrams in Fig. \ref{Fig3} (a) and (b), we obtain the following RG equations in the one-loop approximation:
\begin{eqnarray}
\frac{dg_1}{dy}&=&g_1(g_1+2g_4)+2d_1g_1(g_2-g_1)-2d_3g_1g_2,\nonumber\\
\frac{dg_2}{dy}&=&2(g_1g_4+g_1g_2-g_2g_4-g_2^2)+d_1g_2^2-d_3(g_1^2+g_2^2),\nonumber\\
\frac{dg_4}{dy}&=&g_4^2+2g_1^2-4g_2^2+4g_1g_2-d_4(y)g_4^2.\label{g4A}
\end{eqnarray}
Next, we introduce the test vertices in the same way as in the main text, and derive their renormalization using Feynman diagrams in Fig. \ref{Fig3} (c). For the instabilities due to uniform densities, we find
\begin{eqnarray}
\frac{dn_{\sigma\alpha}}{dy}&=&-g_4n_{\bar{\sigma}\alpha}+\sum_{\beta\neq\alpha}[(g_1-g_2)n_{\sigma\beta}-g_2n_{\bar{\sigma}\beta}],\label{unifA}
\end{eqnarray}
where $\beta$ and $\alpha$ denote the patches belonging to the same valley. We find four distinct eigenvalues corresponding to charge, $\alpha_{c1}$ and $\alpha_{c2}$, antiferromagnetic, $\alpha_{AFM}$, and ferromagnetic, $\alpha_{FM}$, instabilities:
\begin{eqnarray}
\alpha_{c1}&=&2(2G_1-4G_2-G_4)\label{c1A},\\
\alpha_{c2}&=&2(-G_1+2G_2-G_4)\label{c2A},\\
\alpha_{AFM}&=&2(-G_1+G_4)\label{AFMA},\\
\alpha_{FM}&=&2(2G_1+G_4)\label{FMA}.
\end{eqnarray} 

\begin{figure}[h]
\includegraphics[width=0.45\textwidth]{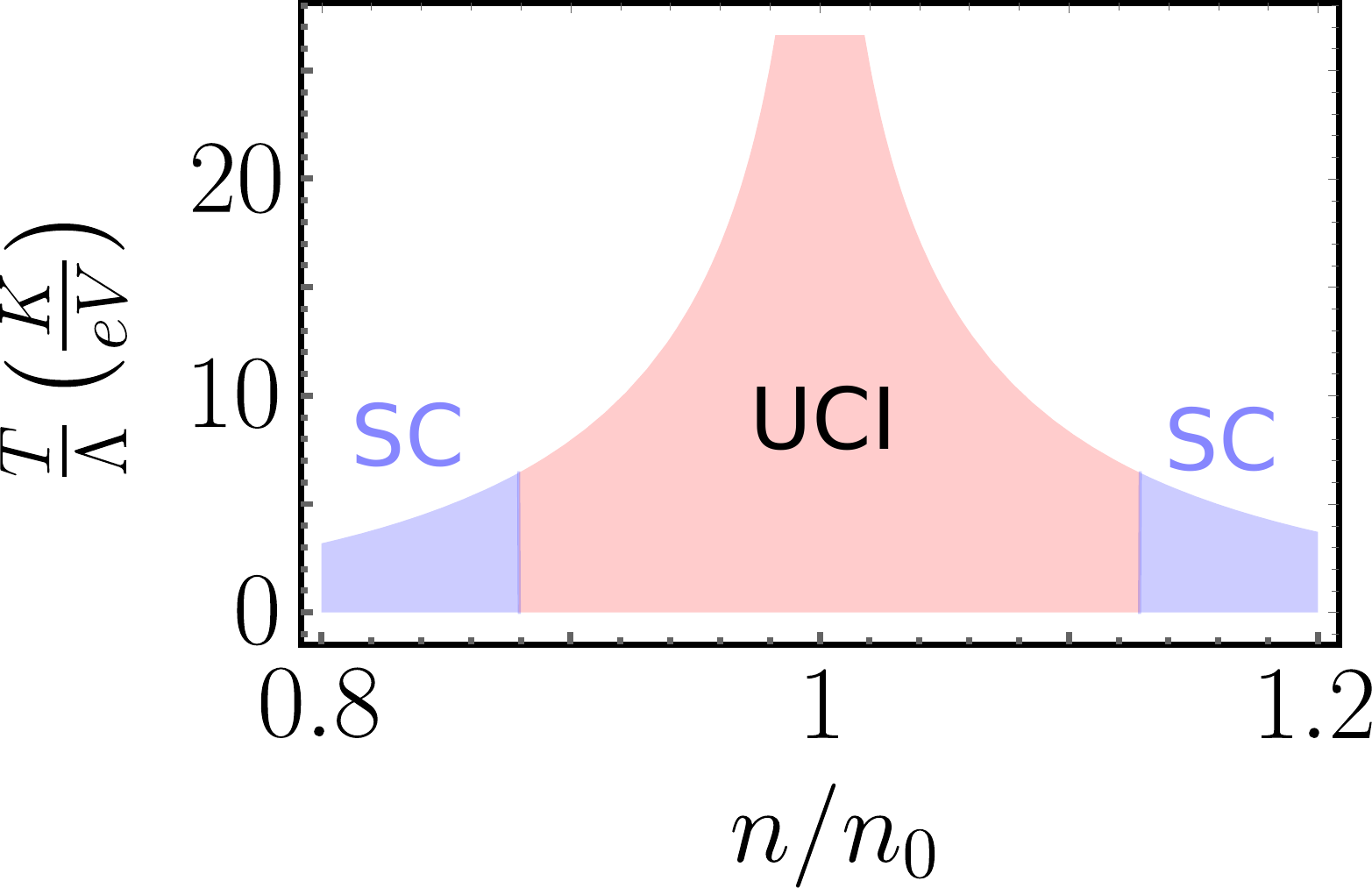}
\caption{Phase diagram for the system with $g_0^{inter}\ll g_0^{intra}$ close to the van Hove filling as a function of doping charge density, $n$, relative to the doping charge density, $n_0$, corresponding to the van Hove filling for the set of parameters: $g_0=0.2$ and  $\epsilon_0=mv^{*2}/2=10^{-2}\Lambda$. The two intra-patch  $s-$wave superconducting phases   are separated by a phase with a uniform charge instability.  }
\label{Fig2A}\end{figure}

For the charge and spin density wave instability, the vertex renormalization is given by Eq. (\ref{nq}), with the charge- and spin-density wave susceptibilities, $\alpha_{CDW}$, and, $\alpha_{SDW}$, given by Eqs. (\ref{CDW1}) and (\ref{SDW1}) respectively.
Finally, the renormalization of the superconducting test vertices are given by Eqs. (\ref{sc1}) and (\ref{sc2}) for intra- and inter-patch superconductivity respectively, leading to the two susceptibilities with $s-$ wave order parameter, $\alpha_{1s}$ and $\alpha_{2s}$, given by Eqs. (\ref{s1}) and (\ref{2s}) respectively, as well as the susceptibility with $s^{\pm}$ order parameter, $\alpha_{s^{\pm}}$, given by Eq. (\ref{2spm}).

The phase diagram for the system with finite $\mu$ as a function of doping charge density shown in Fig. \ref{Fig2A} is similar to the one obtained for the case of valley-independent scattering. However, in the case of negligible inter-valley scattering, the superconducting phase has $s^{++}$ order parameter in contrast with the case of valley-independent scattering, where the order parameter is $s^{\pm}$.

\bibliography{Bilayer_graphene_7}{}
\end{document}